\documentclass[11pt]{article}
\usepackage{amsmath,amssymb}

\baselineskip %
\advance %
\textheight by %
\topskip %
\makeatletter
\@addtoreset{equation}{section}
    
\makeatother

\begin{document}

\title{
\begin{flushright}
{\small CERN-PH-TH/2007-220,\,\,
UB-ECM-PF-07-32,\,\, Toho-CP-0786}\\[1.0cm]
\end{flushright}
{\bf Anisotropic harmonic oscillator, non-commutative Landau
problem and exotic Newton-Hooke symmetry}}

\author
{{\sf Pedro D. Alvarez${}^a$}, {\sf Joaquim Gomis${}^{b,}{}^c$},
{\sf Kiyoshi Kamimura${}^d$}\, {\sf and Mikhail S.
Plyushchay${}^a$}\thanks{E-mails: pd.alvarez.n@gmail.com,
gomis@ecm.ub.es, kamimura@ph.sci.toho.u.ac.jp,
mplyushc@lauca.usach.cl}
\\[4pt]
{\small \it ${}^a$Departamento de F\'{\i}sica, Universidad de
Santiago de Chile, Casilla 307, Santiago 2, Chile}\\
{\small \it ${}^b$Deparment ECM, Facultat de F\'{\i}sica, Universitat de Barcelona, E-08028, Spain}\\
{\small \it ${}^c$PH-TH Division, CERN, CH-1211 Geneva 23, Switzerland}\\
{\small \it ${}^d$Department of Physics, Toho University Funabashi
274-8510, Japan }}
\date{}

\maketitle

\begin{abstract}
We investigate the planar anisotropic harmonic oscillator with
explicit rotational symmetry as a particle model with
non-commutative coordinates. It includes the exotic Newton-Hooke
particle and the non-commutative Landau problem as special,
isotropic and maximally anisotropic,  cases. The system is described
by the same (2+1)-dimensional exotic Newton-Hooke symmetry as in the
isotropic case, and develops three different phases depending on the
values of the two central charges. The special cases of the exotic
Newton-Hooke particle and non-commutative Landau problem are shown
to be characterized by additional, $so(3)$ or $so(2,1)$ Lie
symmetry, which reflects their peculiar spectral properties.
\end{abstract}

\date{}

\section{Introduction}
Classical and quantum theories in 2+1 dimensions possess various
exotic properties. These include, in particular, a possibility for
existence of particles with fractional spin and statistics --
anyons.  Another peculiar property is an equivalence of a
classical (2+1)-dimensional pure gravity to a Chern-Simons gauge
theory.

In a special non-relativistic limit, that is an In\"on\"u-Wigner
contraction, (2+1)D Poincar\'e symmetry of a free anyon theory is
reduced to an exotic Galilei symmetry with two central charges
\cite{LL,BGO,BGGK,LSZ,DH1,JackN,HP1,PHMP}. A similar limit applied
to the AdS$_3$, that is an asymptotic symmetry of the BTZ black
hole solution of the 3D pure gravity \cite{BTZ}, produces an
exotic Newton-Hooke (ENH) symmetry with two central charges
\cite{Mariano,Gao,AGKP}. Both exotic,  Galilei and Newton-Hooke,
symmetries can be realized as symmetries of a particle on a
non-commutative plane. The latter symmetry is transformed into the
former one in a flat limit. The two-fold central extensions of the
Galilei and Newton-Hooke symmetries are possible only in 2+1
dimensions\footnote{The case of an exotic non-relativistic string
in 3+1 dimensions   and the relation with the exotic particle in
2+1 dimensions has been recently studied in \cite{Casal}.}.

Like the BTZ black hole solution \cite{BTZ}, a particle system with
(2+1)D exotic Newton-Hooke symmetry displays three different phases
in dependence on the values of the model parameters \cite{AGKP}.  On
the other hand, its reduced phase space description reveals a
symplectic structure similar to that of Landau problem in the
non-commutative plane \cite{DH1,HL,HP3}. The noncommutative Landau
problem  (NLP) also develops three phases, the sub- and
super-critical ones, separated by a critical, quantum Hall effect
phase \cite{HP3}. Therefore, these similarities indicate on a
possible close relation between the (2+1)D exotic Newton-Hooke
symmetry and the non-commutative Landau problem. The purpose of this
article is to study in detail this relation by means of a planar
exotic \emph{anisotropic} harmonic oscillator with explicit
\emph{spatial rotation} symmetry as a particle model with
non-commutative coordinates.

The model of the anisotropic harmonic oscillator we propose
[(\ref{L_gen}) below], includes the exotic Newton-Hooke particle
and the non-commutative Landau problem as special, isotropic and
maximally anisotropic, cases.  We shaw that what distinguishes the
exotic Newton-Hooke particle and non-commutative Landau problem as
special cases, is a presence of the additional, $so(3)$ or
$so(2,1)$ Lie symmetry.  In a generic case of commensurable
frequencies, the exotic anisotropic harmonic oscillator  is
characterized, instead, by a nonlinear deformation of the
indicated additional Lie symmetry. Like the exotic Newton-Hooke
particle and non-commutative Landau problem, the anisotropic
oscillator system develops the subcritical and supercritical
phases, separated by a critical phase. The phase is defined by the
values of the two central charges of the exotic Newton-Hooke
algebra.

The paper is organized as follows. In Section 2  we introduce a
planar anisotropic harmonic oscillator with explicit rotational
symmetry as a particle model with non-commutative coordinates, and
establish its relation with the non-commutative Landau problem. In
Section 3 we discuss the chiral form of the exotic Newton-Hooke
symmetry of the system, and analyze its additional symmetries, which
depend on the concrete values of the model parameters. In Section 4
we analyze the exotic Newton-Hooke symmetry in the non-chiral,
space-time picture. Section 5 is devoted to the discussion and
concluding remarks.

\section{Planar anisotropic harmonic oscillator and
non-commutative Landau problem}

A canonical Lagrangian of one dimensional harmonic oscillator of
mass $m$ and frequency $\omega$ is given by
\begin{equation}\label{Lharm}
    L_{can}=\frac{\mu}{2}\left(\epsilon_{ij}\dot{X}_iX_j-
    \frac{\alpha}{R}X_i^2\right),
\end{equation}
where $m=\alpha^{-1}\mu R$, $\omega=\alpha R^{-1}$, and $\alpha$
is a dimensionless parameter. Variable $X_1$ can be identified as
the coordinate of one dimensional particle, and then $X_2$ is
proportional to its momentum. Symplectic structure, $
    \{X_i,X_j\}=\frac{1}{\mu}\epsilon_{ij},
$ and Lagrangian (\ref{Lharm}) possess  a two dimensional
\emph{phase space} rotational symmetry. Taking a sum of $n$ copies
of (\ref{Lharm}) with independent parameters $\mu$'s and
$\alpha$'s, we obtain a generalized system of $n$ non-interacting
harmonic oscillators with different frequencies.

Let us consider the case  $n=2$, and take the canonical Lagrangian
in the form
\begin{equation}
    L_{+-}=-\frac{\mu _{+}}{2}\left( \epsilon_{ij}
    \dot{X}_{i}^{+}X_{j}^{+}+ \frac{\alpha_+}{R}{X_{i}^{+}}^{2}\right)
    -\frac{\mu _{-}}{2}\left( -\epsilon_{ij}
    \dot{X}_{i}^{-}X_{j}^{-}+\frac{\alpha_-}{R}{X_{i}^{-}}
    ^{2}\right).
    \label{L_gen}
\end{equation}
We  suppose that $R>0$ and that $\mu_\pm$ can take values of any
sign.  For the moment we do not assume any restrictions for the
parameters $\alpha_\pm$. The dynamics of (\ref{L_gen}) is given by
\begin{equation}
    \dot{X}_{i}^{\pm }\pm \omega_\pm\epsilon _{ij}X_{j}^{\pm }=0.\qquad
    \omega_\pm=\alpha_\pm R^{-1},
    \label{eqs}
\end{equation}
while its symplectic structure is
\begin{equation}
    \{X_i^+,X_j^+\} =- \frac{1}{\mu _+}\epsilon _{ij},\qquad
    \{X_i^-,X_j^-\} = \frac{1}{\mu _-}\epsilon _{ij}, \qquad
    \{X_{i}^+,X_{j}^-\}=0. \label{Xi+-,Xj+-}
\end{equation}
In the chosen special case $n=2$, a \emph{phase space} index $i$
can be reinterpreted as a \emph{spatial} index of the
(2+1)-dimensional space-time. With such reinterpretation,
Lagrangian (\ref{L_gen}) as well as equations of motion
(\ref{eqs}) and symplectic structure (\ref{Xi+-,Xj+-}) possess the
explicit \emph{spatial} $SO(2)$ rotation  symmetry. This
corresponds to the diagonal part of the obvious chiral rotation
symmetry $SO(2)\times SO(2)$ of (\ref{L_gen}). The nondiagonal
part is identified with the time translation symmetry, see Eqs.
(\ref{trans2ch}) and (\ref{hj}) below.

As a result, system (\ref{L_gen}) provides us with a
\emph{rotational invariant} description of the planar
\emph{anisotropic} harmonic oscillator system.

At this point we would like to clarify under which conditions this
planar anisotropic oscillator can be interpreted as a two
dimensional \emph{particle} system with coordinate $x_i$ and
velocity  $v_i$ related by a usual dynamics equation
\begin{equation}\label{xdotv}
    \dot{x}_i=v_i.
\end{equation}
In accordance with (\ref{eqs}), for such a particle system the
variables $X^+_i$ and $X^-_i$ should have a sense of normal, or
chiral modes.

To this aim, we first note that the transformation
$\alpha_\pm\rightarrow -\alpha_\mp$, $\mu_\pm\rightarrow -\mu_\mp$,
$X^+_i\leftrightarrow X^-_i$ does not change  equations of motion
and the Lagrangian, and  it is sufficient to assume that
$(\alpha_++\alpha_-)\geq 0$. If $(\alpha_++\alpha_-)=0$, the two
chiral modes have exactly the same evolution. This case should be
excluded since it does not allow us to introduce $x_i$ and $v_i$
related by (\ref{xdotv}). Taking also into account that $\alpha_\pm$
appear only in the combination $\alpha_\pm/R$, without loss of
generality we can put
\begin{equation}\label{alpha}
    \alpha_\pm(\chi)=\cos\chi(\cos\chi\pm\sin\chi),\qquad
    -\frac{\pi}{2}<\chi<\frac{\pi}{2}.
\end{equation}
With the normalization chosen in (\ref{alpha}), we have\footnote{ In
what follows it will be more convenient to work, however,  in terms
of $\alpha_\pm$, implying the one-parametric representation
(\ref{alpha}).} $0<(\alpha_++\alpha_-)\leq 1$ and $-1\leq
(\alpha_+-\alpha_-)\leq 1$.

Now we can introduce the coordinate and velocity  of the two
dimensional particle system,
\begin{equation}\label{XXxv}
    X^\pm_i=\alpha_{{}_\mp}  x_i\pm R\epsilon_{ij}v_j,
\end{equation}
\begin{equation}\label{xXX}
    x_i=\frac{1}{\alpha_++\alpha_-}\left(X^+_i+X^-_i\right),
    \qquad
    v_i=\frac{1}{R (\alpha_++\alpha_-)}\,
    \epsilon_{ij}\left(\alpha_-X^-_j-\alpha_+X^+_j\right),
\end{equation}
which satisfy  dynamics relation (\ref{xdotv}).
    In a generic case, in correspondence with (\ref{Xi+-,Xj+-}),
 the coordinate $x_i$ describes a \emph{non-commutative} plane,
\begin{equation}\label{xxnon}
    \left\{x_i,x_j\right\}=\frac{1}{(\alpha_++\alpha_-)^2}\,
    \frac{\mu_+-\mu_-}{\mu_+\mu_-}\epsilon_{ij}.
\end{equation}
The components of the coordinate vector $x_i$ are commutative only
when $\mu_+=\mu_-$. As we shall see, only in this case the Galilean
boosts mutually commute. Other brackets are
\begin{equation}\label{xxvv}
    \{x_i,v_j\}=\frac{1}{R^2 (\alpha_++\alpha_-)^2}\,
    \frac{\mathcal{M}}{\mu_+\mu_-}\delta_{ij},\qquad
    \{v_i,v_j\}=\frac{1}{R^2 (\alpha_++\alpha_-)^2 }\,
   \frac{\mathcal{B}}{\mu_+\mu_-} \epsilon_{ij},
\end{equation}
where
\begin{equation}\label{MNma}
    \mathcal{M}=R\left(\mu_+\alpha_- +
    \mu_-\alpha_+\right),\qquad
     \mathcal{B}=\mu_+\alpha_-^2 -\mu_-\alpha_+^2\,.
\end{equation}
Symplectic two-form corresponding to (\ref{xxnon}) and (\ref{xxvv})
has a simple structure,
\begin{equation}\label{2form}
    \sigma={\cal M}dv_i\wedge dx_i+
    \frac{1}{2}R^2(\mu_+-\mu_-)\epsilon_{ij}dv_i\wedge dv_j
    +\frac{1}{2}{\cal B}\epsilon_{ij}dx_i\wedge dx_j.
\end{equation}

 In terms of the particle coordinate $x_i$, the anisotropy
reveals itself in the coupled dynamics of the components $x_1$ and
$x_2$,
\begin{equation}\label{couposc}
    \ddot{x_i}+(\omega_+-\omega_-)
    \epsilon_{ij}\dot{x}_j+\omega_+\omega_-x_i=0,
\end{equation}
where $ \omega_\pm$ are defined in (\ref{eqs}). This should be
compared with  the second order equations for the chiral modes,
$\ddot{X}^\pm_i+\omega_\pm^2 X^\pm_i=0$. The dynamics of $x_1$ and
$x_2$ decouples only in the isotropic ($\chi=0$) case
\begin{equation}\label{aa=1}
    \alpha_+=\alpha_-=1.
\end{equation}

 In terms of variables $x_i$ and
$v_i$, Lagrangian (\ref{L_gen}) takes, up to a total derivative, a
non-chiral form
\begin{equation}
    L =\mathcal{M}\left(\dot{x}_{i}
    v_{i}-\frac{\alpha_+\alpha_-}{2R^2}x_{i}^{2}-
    \frac{\mathcal{N}}{2\mathcal{M}}v_{i}^{2}\right)+\left(
    \mu_{-}-\mu_{+}\right)\left(\alpha_+\alpha_-\,\epsilon_{ij}
    x_{i}v_{j}-\frac{R^{2}}{2}\epsilon _{ij}v_i\dot{v}_{j}\right)+
    \frac{\mathcal{B}}{2}\epsilon _{ij}x_{i}\dot{x}_{j}\, ,
    \label{Lncom}
\end{equation}
where
    $\mathcal{N}=R\left(\mu_+\alpha_+ +
    \mu_-\alpha_-\right)$.
Note  that $\mathcal{M}$ and $\mathcal{N}$ have units of mass,
while $\mathcal{B}$ does of a magnetic field. The term $\epsilon
_{ij}{v}_{i}\dot{v}_{j}$, with coefficient proportional to
$(\mu_+-\mu_-)$ in (\ref{Lncom}) is responsible for the coordinate
non-commutativity (\ref{xxnon}). Physically it describes a
magnetic like coupling for velocities. The terms
$\dot{x}_{i}v_{i}$ and $\epsilon _{ij}x_{i}\dot{x}_{j}$, with
coefficients $\mathcal{M}$ and $\frac{1}{2}\mathcal{B}$,
correspond to the nontrivial brackets between $x_i$ and $v_i$, and
to the velocity non-commutativity, see the first and second
relations in (\ref{xxvv}).

 The equations of motion obtained by variation of (\ref{Lncom}) in
 $x_i$ and $v_i$,
are, respectively,
\begin{eqnarray}
    \mathcal{M}\left(\dot{v}_i+\frac{\alpha_+\alpha_-}{R^2}x_i\right)=
    \epsilon_{ij}\large(\mathcal{B}\dot{x}_j+(\mu_--\mu_+)\alpha_+\alpha_-
    \, v_j\large),\label{ec x_i}\\
    R^2(\mu_+-\mu_-)\left(\dot{v}_i+\frac{\alpha_+\alpha_-}{R^2}x_i\right)=
    \epsilon_{ij}\left(\mathcal{M}\dot{x}_j
    -\mathcal{N}v_j\right).
\label{ecvi}
\end{eqnarray}

If $\mu_+=\mu_-$, and hence, $\left\{x_i,x_j\right\}=0$,
(\ref{ecvi}) takes the form (\ref{xdotv}), that is  an algebraic
equation for $v_i$. In this case $v_1$ and $v_2$  become auxiliary
variables and can be eliminated by their equations of motion  from
(\ref{Lncom}). As a result, (\ref{Lncom}) turns into a regular,
second order Lagrangian $L(x,\dot{x})$, that describes a usual
Landau problem in the presence of additional isotropic harmonic
potential term.

 It is necessary to note that though for $\mu_+\neq \mu_-$
 Eq. (\ref{xdotv})
 also appears as a consequence of the system of equations (\ref{ec x_i}) and
 (\ref{ecvi}), it is not produced by
 the variation of Lagrangian in $v_i$ itself. If
in this case we try to substitute $v_i$, using (\ref{xdotv}), in
Lagrangian
 (\ref{Lncom}),
 we would get a higher derivative, nonequivalent  Lagrangian,
that generates the equations of motion different from
(\ref{couposc}).

In the isotropic case (\ref{aa=1}), system (\ref{Lncom}) is reduced
to the exotic Newton-Hooke particle, which was constructed in
\cite{AGKP} by  the nonlinear realization method \cite{Coleman}
accommodated  to the space-time symmetries, see for example
\cite{GomisK}.

Let us show now that the case of the maximal anisotropy,
\begin{equation}\label{aa0}
    \chi=\varepsilon\frac{\pi}{4}: \qquad \alpha_\varepsilon=1,\quad
    \alpha_{-\varepsilon}=0,\qquad
    \varepsilon=+,-,
\end{equation}
corresponds to the non-commutative Landau problem  described by the
Hamiltonian
\begin{equation}
    H=\frac{1}{2m}{\mathcal{P}}_{i}^{2},
    \label{Hnclp}
\end{equation}
symplectic structure
\begin{equation}
    \left\{ {\mathcal{X}}_{i},{\mathcal{X}}_{j}\right\} =\frac{\theta}
    {1-\beta } \epsilon _{ij},\qquad \left\{
    {\mathcal{X}}_{i},{\mathcal{P}}_{i}\right\} = \frac{1}{1-\beta}
    \delta _{ij},\qquad \left\{ {\mathcal{P}}_{i},{\mathcal{P}}_{j}
    \right\} =\frac{B}{1-\beta }\epsilon _{ij},
    \label{Poissonnclp}
\end{equation}
and equations of motion
\begin{equation}
    \dot{\mathcal{X}}_{i}=\frac{1}{m^*}\mathcal{P}_{i},\qquad
    \dot{\mathcal{P}}_{i}=\frac{B}{m^*}\epsilon_{ij}\mathcal{P}_{j}.
    \label{ecmovnclp}
\end{equation}
Here $B$  is magnetic field, $\beta=\theta B$, $m^*=m(1-\beta)$
plays the role of the effective mass, while $\theta$ is a parameter,
which at $B=0$ characterizes a non-commutativity of the coordinates
and of the Galilean boosts of a free exotic particle \cite{DH1,HP3}.
Lagrangian  corresponding to (\ref{Hnclp}) and (\ref{Poissonnclp})
is given by
\begin{equation}\label{NLPlag}
    L_{{NLP}}={\cal P}_i\dot{\cal X}_i -\frac{1}{2m}{\cal P}_i^2
    +\frac{1}{2}\theta\epsilon_{ij}{\cal P}_i\dot{\cal P}_j+
    \frac{1}{2}B\epsilon_{ij}{\cal X}_i\dot{\cal X}_j\,.
\end{equation}

Comparing (\ref{Lncom}) and (\ref{NLPlag}), we find that in
maximally anisotropic case (\ref{aa0}) the former system reduces to
the latter  one under the following correspondence between the
variables and parameters:
\begin{equation}\label{xXvP}
    x_i={\cal X}_i,\qquad v_i=\frac{1}{m^*}{\cal P}_i,
\end{equation}
\begin{equation}
    \mu_{\varepsilon}=\vert B(1-\theta B)\vert,\qquad
      \mu_{-\varepsilon}=\vert B\vert {\rm sgn}(1-\theta B),\qquad
    R=\left\vert\omega^{-1}\right\vert,\label{ahoNLP}
\end{equation}
\begin{equation}
    \varepsilon={\rm sgn} \large(B(\beta-1)\large),\qquad
       \omega=\frac{B}{m^*}.
  \label{omega}
\end{equation}

 Having this correspondence, and using transformation
(\ref{xXX}), we obtain the chiral form of Lagrangian (\ref{NLPlag})
for the case $\alpha_+=0$, $\alpha_-=1$ ($\varepsilon=-1$) ,
\begin{equation}
    L_{NLP}^{+-}=-\frac{B}{2} \epsilon_{ij}
    \dot{X}_{i}^{+}X_{j}^{+} -\frac{B(1-\beta)}{2}\left(
    -\epsilon _{ij}\dot{X}_{i}^{-}X_{j}^{-}+
    \omega{X_{i}^{-}}^{2}\right),
    \label{NLPchir}
\end{equation}
which generates equations of motion
\begin{equation}
    \dot{X}_{i}^{+}=0, \qquad \dot{X}_{i}^{-}-\omega
    \epsilon_{ij}X_{j}^{-}=0.
    \label{NLPeqX}
\end{equation}
In terms of the variables ${\cal X}_i$ and ${\cal P}_i$, the
chiral (normal) modes are given by
\begin{equation}
    X_{i}^{+}={\cal X}_i+\frac{1}{m^*|\omega|}\epsilon_{ij}{\cal P}_j, \qquad
    X_{i}^{-}=-\frac{1}{m^*|\omega|}\epsilon_{ij}{\cal P}_j.
\end{equation}
The chiral mode $X^+_i$ is an integral of motion not depending
explicitly on time (cf. the chiral integrals of motion
(\ref{calJi+-}) in a generic case). It can be identified as a
guiding center coordinate.

The case $\alpha_+=1$, $\alpha_-=0$ ($\varepsilon=+1$) can be
obtained via obvious changes in correspondence with relations
(\ref{ahoNLP}), (\ref{omega}).  In this case the chiral mode $X^-_i$
plays the role of the guiding center coordinate, while $X^+_i$ has
the same evolution law as the chiral mode $X^-_i$ in the previous
case $\varepsilon=-1$.

The flat limit
\begin{equation}
    R\rightarrow\infty,\qquad (\mu_+-\mu_-)\rightarrow 0,
    \qquad
     R\mu_+(\alpha_++\alpha_-)\rightarrow m,\qquad
     R^2(\mu_+-\mu_-)\rightarrow \theta m^2,\label{freeEx}
\end{equation}
applied to (\ref{Lncom}), produces a free exotic particle,
\begin{equation}\label{Lfree}
    L_\theta=m\left(\dot{x}_iv_i-\frac{1}{2}v_i^2\right)+\frac{1}{2}\theta
    m^2\epsilon_{ij}v_i\dot{v}_j,
\end{equation}
that is described by the equations of motion $\dot{x}_i=v_i$,
$\dot{v}_i=0$,  and carries the two-fold centrally extended Galilei
symmetry \cite{DH1}. If, as in  generic case (\ref{Lncom}), we try
to substitute $v_i$ using the equation $v_i=\dot{x}_i$ produced by
variation of (\ref{Lfree}) in $x_i$, we would get a nonequivalent
higher derivative model \cite{LSZ} with additional, spin degrees of
freedom, see \cite{HP1,PHMP}.

\section{Symmetries: chiral picture}

To identify the symmetries of our system, we proceed from the chiral
Lagrangian (\ref{L_gen}). We integrate equations (\ref{eqs}),
\begin{equation}\label{Xt}
    X_i^{\pm}(t)=\Delta_{ij}^{\pm}(t)X_j^{\pm}(0), \qquad
    \Delta_{ij}^{\pm}(t)=\delta_{ij}\cos(\alpha_{\pm}t/R)\mp\epsilon_{ij}\sin(\alpha_{\pm}t/R),
\end{equation}
and construct the integrals of motion,
\begin{equation}
    \mathcal{J}_{i}^{\pm }\equiv
    R\mu_{\pm}\epsilon_{ij}X_{j}^{\pm}(0)=R\mu_{\pm}\epsilon_{ij}
    \Delta_{jk}^{\pm}(-t)X_{k}^{\pm},
    \label{calJi+-}
\end{equation}
\begin{equation}
    \mathcal{J}^{\pm }\equiv
    \pm\frac{\mu_{\pm}}{2}\left(X_{i}^{\pm}(0)\right)^{2}
    =\pm\frac{\mu_{\pm}}{2}\left(X_{i}^{\pm}\right)^{2},
    \label{calJ+-}
\end{equation}
where $X^\pm_i=X^\pm_i(t)$. The quantities $\mathcal{J}_{i}^{\pm}$
are the integrals of motion that include explicit dependence on
time and  satisfy the equation $\frac{d}{dt}\mathcal{J}_{i}^{\pm
}= \frac{\partial}{\partial t}\mathcal{J}_{i}^{\pm
}+\{\mathcal{J}_{i}^{\pm },H\}=0$, where
\begin{equation}\label{Hamil}
    H=\frac{1}{2R}\left(\mu_+\alpha_+X_i^+{}^2+\mu_-\alpha_-X_i^-{}^2\right)
\end{equation}
plays the role of the Hamiltonian. Unlike the linear in $X^\pm_i$
integrals (\ref{calJi+-}), the quadratic integrals (\ref{calJ+-})
do not include explicit dependence on time.

Integrals (\ref{calJi+-}) and (\ref{calJ+-}) generate the algebra
\begin{equation}
    \{\mathcal{J}^{+},\mathcal{J}_{i}^{+}\}=\epsilon _{ij}\mathcal{J}_{j}^{+},
    \qquad
    \{\mathcal{J}_{i}^{+},\mathcal{J}_{j}^{+}\}=Z^{+}\epsilon _{ij},
    \label{chiral+}
\end{equation}
\begin{equation}
    \{\mathcal{J}^{-},\mathcal{J}_{i}^{-}\}=\epsilon _{ij}\mathcal{J}_{j}^{-},
    \qquad
    \{\mathcal{J}_{i}^{-},\mathcal{J}_{j}^{-}\}=Z^{-}\epsilon _{ij},
    \label{chiral-}
\end{equation}
where $Z^\pm=\mp R^2\mu_{\pm}$ have a sense of central charges, and
all other brackets are equal to zero. This is a chiral form of the
(2+1)D exotic Newton-Hooke symmetry presented in the form of a
direct sum of the two (1+1)D centrally extended Newton-Hooke
algebras. The quadratic Casimirs of this algebra are
\begin{equation}\label{Casimir}
    {\cal C}_\pm= {{\cal J}^\pm_i}^2 + 2Z^\pm {\cal J}^\pm \, .
\end{equation}

Let us stress that the exotic Newton-Hooke algebra (\ref{chiral+}),
(\ref{chiral-}) in a generic case (\ref{alpha}) has exactly the same
form as in the particular isotropic case (\ref{aa=1}). For the
latter case it was obtained in \cite{AGKP} by a contraction of the
AdS${}_3$ algebra, with identification of the parameter $R$ as the
AdS${}_3$ radius. This chiral form of the ENH symmetry is rooted in
the algebra isomorphism
$$
AdS_3\sim so(2,2)\sim so(2,1)\oplus so(2,1)\sim AdS_2\oplus AdS_2.
$$

Integrals (\ref{calJi+-}) and (\ref{calJ+-}) generate the symmetry
transformations of the chiral coordinates\footnote{We do not
indicate explicitly infinitesimal transformation parameters.},
\begin{eqnarray}
   & \{ X_{i}^{+ },\mathcal{J}_{j}^{+ }\} =- R\Delta_{ij}^{+}(t), \qquad
    \{X_{i}^{- },\mathcal{J}_{j}^{- }\} = R\Delta_{ij}^{-}(t),&
\label{trans1ch}\\
    &\{ X_{i}^{+ },\mathcal{J}^{+}\} = -\epsilon _{ij}X_{j}^{+},\qquad \{
    X_{i}^{- },\mathcal{J}^{-}\} =-\epsilon_{ij}
    X_{j}^{-}.&\label{trans2ch}
\end{eqnarray}
Due to the presence of the explicit dependence on time of the
chiral integrals (\ref{calJi+-}), symmetry transformations
(\ref{trans1ch}) are also time-dependent. Under them, Lagrangian
(\ref{L_gen}) is quasi-invariant.

The Hamiltonian  and the angular momentum are identified as linear
combinations of the quadratic integrals, which generate  the time
translations and space rotations,
\begin{equation}
    H=\frac{1}{R}\left( \alpha _{+}\mathcal{J}^{+}-\alpha_{-}
    \mathcal{J} ^{-}\right) ,\qquad
    J=\mathcal{J}^{+}+\mathcal{J}^{-}.
    \label{hj}
\end{equation}
This form of $H$ and $J$ is behind the anisotropy of dynamics and
rotational symmetry of  the system (\ref{L_gen}).

For the NLP case  (\ref{aa0}), one of the chiral generators,
${\cal J}^+$ or ${\cal J}^-$, disappears from the Hamiltonian. As
a result, the corresponding  chiral mode  has a trivial dynamics
of the guiding center coordinate, $\dot{X}^+=0$, or
$\dot{X}^-_i=0$, see Eq. (\ref{NLPeqX}), and one of the two
vectors (\ref{calJi+-}) transforms into an integral of motion that
does not depend explicitly on time.

Notice that in the excluded case $\alpha_+=-\alpha_-$, the time
translation, $H$,  and rotation, $J$, generators would be (up to a
multiplicative constant) the same, that prevents the introduction of
the coordinate $x_i$ and velocity $v_i$ related by Eq.
(\ref{xdotv}). \vskip0.2cm

Let us discuss briefly symmetries  in the quantum case.

We define the operators
\begin{equation}
    a^{-}=\sqrt{\frac{\left\vert \mu _{+}\right\vert }{2}}\left(
    X_{2}^{+}+iX_{1}^{+}\right) ,\quad a^{+}=\left( a^{-}\right) ^{\dag},
    \qquad b^{-}=\sqrt{\frac{\left\vert \mu _{-}\right\vert}{2}}
    \left( X_{1}^{-}+iX_{2}^{-}\right) ,\quad b^{+}=\left(b^{-}\right) ^{\dag },
\end{equation}
which obey the commutation relations $\left[ a^{-},a^{+}\right]
=\epsilon_+$, $\left[ b^{-},b^{+}\right] =\epsilon_-$ and
$[a^\pm,b^\pm]=0$, where $\epsilon_\pm={\rm sgn} (\mu_\pm)$.  For
$\mu_+>0$ ($\mu_+<0$) and $\mu_->0$ ($\mu_-<0$), the operators
$a^+$ ($a^-$) and $b^+$ ($b^-$) are identified as creation
oscillator operators. With the symmetrized ordering prescription,
(\ref{calJ+-}) and (\ref{hj}) give the Hamiltonian and the angular
momentum operators in a form
\begin{equation}
    RH= \epsilon_+\alpha_{+}\,a^{+}a^{-}+
    \epsilon_-\alpha_{-}\,b^{+}b^{-}
    +\frac{1}{2}\left(\epsilon_+\alpha_{+}+
    \epsilon_-\alpha_{-}\right),
\end{equation}
\begin{equation}
    J=\epsilon_+\,a^{+}a^{-}-
    \epsilon_-\,b^{+}b^{-}+\frac{1}{2}\left(\epsilon_+-
    \epsilon_-\right).
\end{equation}
{}From here it follows that in  the generic anisotropic case, like
in particular cases of  the ENH  particle \cite{AGKP} and the NLP
\cite{HP3}, our system can exhibit three different kinds of behavior
in dependence on the values of the parameters $\mu_\pm$.

Since the eigenvalues of the number operators take non-negative
integer values, $n_a$, $n_b=0,1,...$, we find that when
$\epsilon_+=\epsilon_-$, $J$ can take values of both signs. Its
spectrum is unbounded. This case we call a \emph{subcritical}
phase. The spectrum of $H$ in this phase is bounded from below
(when $\mu_+,\mu_->0$), or from above (for $\mu_+,\mu_-<0$).

When the signs of $\mu_+$ and $\mu_-$ are opposite, the spectrum of
$J$ is bounded from one side. This is a \emph{supercritical} phase.
In this phase the spectrum of $H$ is unbounded, except the NLP case.
The case of the NLP is special: its energy is bounded in both, sub-
and super-critical  phases, because one of the chiral modes has zero
frequency, and does not contribute to the energy.

Yet, another phase corresponds to the case when one of the
parameters $\mu_+$ or $\mu_-$ takes zero value. In such a phase one
of the modes disappears from Lagrangian (taking a role of a pure
gauge degree of freedom), and the system  transforms into a
one-dimensional oscillator, whose symmetry is described by the
(1+1)D centrally extended Newton-Hooke algebra \cite{AGKP}. This is
a \emph{critical} phase that separates two other phases, and is
characterized by the zero value of one of the central charges,
$Z^+$, or $Z^-$. In the NLP it corresponds to the quantum Hall
effect phase. We note here that in the critical phase, in turn, two
different cases should be distinguished. When, say, $\mu_-=0$ and
$\alpha_+=1$, the Hamiltonian is nontrivial and generates a rotation
of the remaining chiral mode $X^+_i$, that coincides (up to a gauge
shift) with $x_i$. When $\mu_-=0$ and $\alpha_+=0$, Hamiltonian is
equal to zero, and the chiral mode has a trivial dynamics,
$X^+_i(t)=X^+_i(0)$ [for a discussion of the critical phase in the
NLP, see ref. \cite{HP3}]. In both cases, $\alpha_+=1$ and
$\alpha_+=0$, $x_i$ and $v_i$ satisfy relation (\ref{xdotv}), but
they are linearly dependent variables in correspondence with
decreasing of the number of the physical degrees of freedom.

{}From the point of view of the dynamics and symmetries, as in the
case of the usual planar anisotropic harmonic oscillator given by
the second order Lagrangian, it is also necessary to distinguish
special cases. As it follows from (\ref{xXX}) and (\ref{Xt}), the
particle trajectory $x_i(t)$ is closed only when $\alpha_+/\alpha_-$
is rational. Behind this property, there is additional symmetry.

In the isotropic case (\ref{aa=1}), the ENH particle system is
characterized by additional symmetry associated with the integrals
$a^+b^-$ and $a^-b^+$, which, like $H$ and $J$, do not include
explicit dependence on the time. In terms of the chiral modes, these
are linear combinations of the Hermitian operators
$(X^+_1X^-_2+X^+_2X^-_1)$ and $(X^+_2X^-_2-X^+_1X^-_1)$. In sub- and
super- critical phases, they together with angular momentum $J$
generate the $so(3)$ and $so(2,1)$ symmetries, which are responsible
for a finite and infinite degeneracy of the energy levels. This
additional symmetry was discussed in detail in \cite{AGKP}.

In the NLP, analogously, we have additional $so(2,1)$ symmetry. If,
say, $\alpha_+=0$, $\alpha_-=1$, and $\mu_+>0$, the $so(2,1)$
generators are given by the quadratic integrals
$I_0=\frac{1}{4}\{a^+,a^-\}$, $I_+=\frac{1}{2}a^+{}^2$ and
$I_-=\frac{1}{2}a^-{}^2$,
\begin{equation}
    [I_0,I_\pm]=\pm I_\pm,\qquad
    [I_+,I_-]=-2I_0.
    \label{III}
\end{equation}
All the energy levels are infinitely degenerate.

In the anisotropic case with $\alpha_+/\alpha_-=p/q$,
$p,q=1,2,\ldots$, $p\neq q$, the frequencies of the chiral modes
are commensurable, and the system has additional integrals of
motion $j_+=(a^+)^q(b^-)^p$ and $j_-=(a^-)^q(b^+)^p$. In this case
the energy levels have finite, or infinite additional degeneracy,
in dependence on whether we have a sub-, or super- critical phase.
These integrals together with the angular momentum generate a
nonlinear (polynomial) deformation of the $so(3)$, or $so(2,1)$
algebra,
\begin{equation}
    [J,j_\pm]=\pm (q+p) j_\pm\,, \label{Jjj}
\end{equation}
\begin{equation}\label{jjJH}
    [j_+,j_-]=\prod_{k=1}^{q}[a^+a^-
    +(1-k)\epsilon_+]
                \prod_{\ell=1}^{p}[b^+b^-+\ell\epsilon_-]-
\prod_{k=1}^{q}[a^+a^-+k\epsilon_+]
                \prod_{\ell=1}^{p}[b^+b^-+(1-\ell)\epsilon_-]\,,
\end{equation}
where
\begin{equation}
  a^+a^- =\frac{\epsilon_+}{(\alpha_++\alpha_-)}\left(RH+\alpha_-J\right)-
\frac{1}{2},\qquad
  b^+b^- =\frac{\epsilon_-}{(\alpha_++\alpha_-)}\left(RH-\alpha_+J\right)-
\frac{1}{2},
\end{equation}
in which the Hamiltonian plays a role of the central element,
$[H,J]=[H,j_\pm]=0$. This is completely analogous to the very well
known property of a usual (non-exotic) planar anisotropic
oscillator with commensurable frequencies, see
\cite{Jauch,Louck,Boer,Kij}.

\section{Exotic Newton-Hooke  symmetry: space-time picture}

Here we find the symmetry transformations in terms of the
variables $x_i$, and $v_i$ of the (2+1)D particle system, and
their corresponding generators. In particular, we identify the
integrals, which in a flat limit are transformed into commuting
translations and non-commuting boosts generators, and find the
algebra formed by them together with  $H$ and $J$.

To identify the translations and boosts generators, we note that
because of the vector nature, they have to be linear combinations of
the integrals $\mathcal{J}_i^{\pm}$. The transformations produced by
$\mathcal{J}_i^{\pm}$,  in correspondence with relations
(\ref{trans1ch}) and (\ref{xXX}), are
\begin{equation}
    \{
    x_{i},\mathcal{J}_{j}^{\pm}\}=\mp R
    (\alpha_++\alpha_-)^{-1}\Delta_{ij}^{\pm}(t),\qquad
    \{v_{i},\mathcal{J}_{j}^{\pm}\}=\alpha_\pm
     (\alpha_++\alpha_-)^{-1}\epsilon_{ik}\Delta_{kj}^{\pm}(t),
    \label{xvJJ}
\end{equation}
where $\Delta_{ij}^{\pm}(t)$ is defined in (\ref{Xt}). Then, in
order to recover the Galilean transformations in the flat limit
(\ref{freeEx}), $
    \{ x_{i},P_{j}\}=\delta_{ij},
$
$ \{x_{i},K_{j}
    \}=-\delta_{ij}t,
$ $\{ v_{i},P_{j} \}=0,$ $\{v_{i},K_{j} \}=-\delta_{ij},$
we obtain
\begin{equation}
    P_{i}=\frac{1}{R}\left(\alpha_+\mathcal{J}_{i}^{-}-\alpha_-
    \mathcal{J}_{i}^{+}\right),
    \qquad K_{i}=-\epsilon_{ij}\left(\mathcal{J}_{j}^{+}+
    \mathcal{J}_{j}^{-}\right). \label{PK}
    \end{equation}
Note that the relation (\ref{PK}) between $P_i$, $K_i$ and ${\cal
J}^\pm_i$ has a structure similar to that between $H$, $J$ and
${\cal J}^\pm$, see (\ref{hj}). The symmetry transformations
generated by $P_i$ and $K_i$ can be computed by means of
(\ref{xvJJ}) and (\ref{PK}).

 In correspondence with (\ref{xdotv})
and (\ref{couposc}), the time translations symmetry transformations
take here a form
\begin{equation}\label{xvH}
    \{x_i,H\}=v_i,\qquad
    \{v_i,H\}=-\frac{\alpha_+\alpha_-}{R^2}
    x_i-\frac{\alpha_+-\alpha_-}{R}\epsilon_{ij}v_j,
\end{equation}
where, in correspondence with (\ref{Lncom}),
\begin{equation}\label{Hxv}
    H=\alpha_+\alpha_-\left(\frac{\mathcal{M}}{2R^2}x_i^2+
    (\mu_+-\mu_-)\epsilon_{ij}x_iv_j\right)
    +\frac{\cal N}{2}v_i^2.
\end{equation}
 The second term in the transformation law for the velocity
is proportional to the rotation symmetry transformation
$\{v_i,J\}=-\epsilon_{ij}v_j$, and disappears in the isotropic case.
It is worth  to note that the structure of the angular momentum,
\begin{equation}\label{Jxv}
    J=\frac{{\cal B}}{2}x_i^2 +\frac{1}{2}R^2(\mu_+-\mu_-)v_i^2+
    {\cal M}\epsilon_{ij}x_iv_j,
\end{equation}
reproduces the structure of the symplectic two-form (\ref{2form}).

The symmetry algebra generated by $H$, $J$, $P_i$ and $K_i$ is
\begin{equation}\label{KPK}
    \left\{ K_{i},K_{j}\right\} =-\tilde{Z}\epsilon _{ij},\quad
     \left\{ P_{i},P_{j}\right\} =- \frac{1}{R^{2}}\left( R\left(
    \alpha _{+}-\alpha _{-}\right) Z+\alpha _{+}\alpha_{-}
    \tilde{Z}\right) \epsilon _{ij},
    \quad
    \left\{
    K_{i},P_{j}\right\} =Z\delta _{ij},
\end{equation}
\begin{equation}\label{KPJ}
    \left\{ K_{i},J\right\} =-\epsilon _{ij}K_{j},\qquad \left\{ P_{i},J\right\}
    =-\epsilon_{ij}P_{j,}\qquad \left\{ H,J\right\} =0\, ,
\end{equation}
\begin{equation}\label{HJPK}
    \left\{ K_{i},H\right\} =P_{i}+\frac{\left( \alpha _{+}-\alpha _{-}\right) }{R}
    \epsilon_{ij}K_{j},\qquad \left\{ P_{i},H\right\} =-\frac{\alpha _{+}
    \alpha _{-}}{R^{2}}K_{i}\, ,
\end{equation}
where
\begin{equation}\label{ZZZ}
    Z=(\alpha_+Z^--\alpha_-Z^+)R^{-1}={\cal M},\qquad
    \tilde{Z}=-(Z^++Z^-)=R^2(\mu_+-\mu_-),
\end{equation}
and ${\cal M}$ is defined in (\ref{MNma}). Casimirs (\ref{Casimir})
take here an equivalent form
\begin{equation}\label{Cas2}
    \mathcal{C}_{\pm}=\left(P_i \mp
    \frac{\alpha_{\pm}}{R}\epsilon_{ij}K_j\right)^2
    -2\left(Z \pm
    \frac{\alpha_{\pm}}{R}\tilde{Z}\right) \left(H \pm
    \frac{\alpha_{\mp}}{R}J\right).
\end{equation}

When $\mu_+=\mu_-$, the central charge $\tilde{Z}$ takes zero value,
and the Galilean boosts commute. It is exactly the same case when
the coordinates of the particle are commutative, see (\ref{xxnon}).
Analogously, the commutativity of the boosts and translations takes
place when another central charge disappears, $Z={\cal M}=0$. In
this case the coordinate $x_i$ and velocity $v_i$ commute, see the
first relation in (\ref{xxvv}).

For the particular case (\ref{aa0})  of the non-commutative Landau
problem the explicit form of the algebra generated by $H$, $J$,
$P_i$ and $K_i$  can be obtained from (\ref{KPK})--(\ref{Cas2}) by
means of the correspondence relations (\ref{xXvP})--(\ref{omega}).
We only note that the translation generator is reduced here for
the conserved chiral mode identified with the guiding center
coordinate, see (\ref{PK}). It generates usual time-independent
translations, $\delta x_i=\delta a_i$, $\delta v_i=0$, under which
Lagrangian (\ref{NLPlag}) is quasi-invariant.

In the generic case, the generators $H$, $J$, $P_i$ and $K_i$, and
the central charges $Z$ and $\tilde{Z}$ are linear combinations of
the chiral integrals ${\cal J}^\pm$ and ${\cal J}^\pm_i$ and central
charges $Z^+$ and $Z^-$, see Eqs. (\ref{hj}), (\ref{PK}) and
(\ref{ZZZ}). Hence, there exists a linear relation between the
space-time, non-chiral symmetry generators of the exotic anisotropic
harmonic oscillator characterized by the parameters
$\alpha_\pm(\chi)$, and the space-time generators of the exotic
Newton-Hooke symmetry which corresponds to the symmetric case
(\ref{aa=1}). Explicitly, we have
\begin{equation}
    \binom{RH}{J}_{\alpha_+,\alpha_-}=
    \begin{pmatrix}
    {\cal A}_+  & {\cal A}_-  \\
    0 & 1
    \end{pmatrix}
    \binom{RH}{J}_{\alpha_+=\alpha_-=1},
    \label{A1}
\end{equation}
\begin{equation}
    \binom{RP_{i}}{K_{i}}_{\alpha_+,\alpha_-}=
    \begin{pmatrix}
    {\cal A}_+ \delta _{ij} &
    {\cal A}_- \epsilon _{ij} \\
    0 & \delta _{ij}
    \end{pmatrix}
    \binom{RP_{j}}{K_{j}}_{\alpha_+=\alpha_-=1},
    \label{A2}
\end{equation}
\begin{equation}
    \binom{RZ}{\tilde{Z}}_{\alpha_+,\alpha_-}=
    \begin{pmatrix}
    {\cal A}_+  & -{\cal A}_-  \\
    0 & 1
    \end{pmatrix}
    \binom{RZ}{\tilde{Z}}_{\alpha_+=\alpha_-=1},
    \label{A3}
\end{equation}
where
$
     {\cal A}_\pm=\frac{1}{2}(\alpha_+\pm\alpha_-).
$

These  relations mean that the general case of the exotic
anisotropic harmonic oscillator, including the NLP system as a
particular case, is described, in fact, by the same non-chiral,
space-time form of the (2+1)D exotic Newton-Hooke symmetry, as the
exotic isotropic oscillator does. On the other hand, with making use
of  (\ref{A1})--(\ref{A3}), the generators and central charges of
the generic anisotropic case can also be presented as linear
combinations of the generators and central charges of the
non-commutative Landau problem.

\section{Discussion and concluding remarks}

We have showed that the planar exotic anisotropic harmonic
oscillator is characterized by exactly  the same chiral form of
the (2+1)D exotic Newton-Hooke symmetry algebra (\ref{chiral+}),
(\ref{chiral-}) as in the isotropic case. The anisotropy reveals
itself only in the symmetry transformations law, see Eqs.
(\ref{trans1ch}) and (\ref{Xt}), that is associated with the
anisotropy of the Hamiltonian structure (\ref{hj}). In the
space-time  picture the anisotropy reveals itself both in the
symmetry transformations, and in the structure of the symmetry
algebra. However, unlike the usual planar anisotropic harmonic
oscillator given by the second order Lagrangian, anisotropy does
not violate the rotation symmetry.  It only mixes the time
translations with the rotations, and the space translations with
the boosts transformations, see Eqs. (\ref{A1}), (\ref{A2}). The
presence of the two independent parameters $\mu_+$ and $\mu_-$  is
behind the non-commutativity nature of the particle coordinate
(\ref{xxnon}) in a generic case of the model.

The translations, $P_i$, and the boosts, $K_i$, generators of the
exotic Newton-Hooke symmetry are the linear combinations of the
first order in the chiral variables $X^\pm_i$ integrals
(\ref{calJi+-}), that include explicit dependence on time. On the
contrary, the $H$ and $J$ are the linear combinations of the
explicitly time-independent, quadratic in (\ref{calJi+-}),
integrals (\ref{calJ+-}). The additional symmetries of the
isotropic and the NLP special cases, discussed at the end of
Section 3, are also generated by the explicitly time-independent
integrals, which are quadratic in the integrals ${\cal J}^\pm_i$.
If we supply these four explicitly time-independent quadratic
integrals with other six, explicitly  time-dependent quadratic in
${\cal J}^\pm_i$ integrals of motion, we would get a more broad,
the $AdS_4\sim so(3,2)\sim sp(4)$ symmetry as a symmetry of the
system. With respect to the ten $so(3,2)$ generators, which are
certain linear combinations of the quadratic quantities $L_aL_b$,
$a=1,\ldots, 4$, $L_a=({\cal J}^+_i,{\cal J}^-_j)$, the integrals
${\cal J}^\pm_i$ form a Majorana spinor \cite{Sudar,HPV}. {}From
the point of view of the $so(3,2)$ algebra, the Hamiltonian $H$
and the angular momentum $J$ of the exotic anisotropic harmonic
oscillator are just linearly independent combinations of the one
of spatial rotation generators in an abstract (3+2)D space-time,
and of the generator of rotations in the plane of the two
time-like coordinates in that space-time, see \cite{HPV}. We
notice here that the system (\ref{L_gen}) can be related to the
gauge fixed version of the
 $Sp(4)$ gauge invariant particle mechanics model
 \cite{GomisKamimura}.

To conclude, since there are some indications on the possible
close relation of the ENH particle to the physics of the BTZ black
hole \cite{AGKP}, it would be interesting to clarify whether the
exotic anisotropic harmonic oscillator, and the non-commutative
Landau problem as its particular  case, could be related to the 3D
gravity physics. A close relation between the usual Landau problem
and a family of G\"odel-type solutions in M-theory and 3+1 General
Relativity was pointed out recently in \cite{Hik,Druk}.

\vskip 0.4cm\noindent {\bf Acknowledgements}. This work has been
supported in part by
  CONICYT, FONDECYT
Project 1050001 (Chile), the European EC-RTN project
MRTN-CT-2004-005104, MCYT FPA 2004-04582-C02-01 and CIRIT GC
2005SGR-00564. PA thanks the Physics Department of Barcelona
University, where a part of this work was realized, for
hospitality.

\end{document}